\documentclass[letterpaper, 10 pt, conference]{ieeeconf}  % Comment this line out if you need a4paper

\IEEEoverridecommandlockouts                              % This command is only needed if 
\usepackage{algcompatible}
\usepackage{algorithm}
\usepackage[noend]{algpseudocode}
\usepackage{amsmath}
\usepackage{amssymb}
\usepackage{array}
\usepackage[english]{babel}
\usepackage{bigstrut}
\usepackage{bm}
\usepackage{cancel}
\usepackage{caption}
\usepackage{comment}
\usepackage{hyperref}
\usepackage{cleveref}
\usepackage{enumerate}
\usepackage{float}
\usepackage{graphicx}
\usepackage[utf8]{inputenc}
\usepackage{listings}
\usepackage{longtable}
\usepackage{mathrsfs}
\usepackage{multicol}
\usepackage{multirow}
\usepackage{subcaption}
\usepackage{esvect}
\usepackage{upgreek}
\usepackage{color}

\renewcommand{\mod}{\text{mod}}

\renewcommand{\d}{\mathfrak{d}}

\newtheorem{lemma}{Lemma}
\newtheorem{thm}{Theorem}

\newtheorem{remark}{Remark}

\newcommand{\RomanNumeralCaps}[1]
    {\MakeUppercase{\romannumeral #1}}

\makeatletter
\let\sv@thm\@thm
\def\@thm{\let\indent\relax\sv@thm}
\makeatother

\title{\LARGE \bf
Modification of Hilbert's Space-Filling Curve to Avoid Obstacles: \\ A Robotic Path-Planning Strategy
}

\author{Anant A. Joshi$^{1}$, Maulik C. Bhatt$^{2}$ and Arpita Sinha$^{3}$% <-this % stops a space
\thanks{$^{1}$Undergraduate Student, Mechanical Engineering; $^{2}$Undergraduate Student, Aerospace Engineering; $^{3}$Associate Professor, Systems and Control Engineering; Indian Institute of Technology Bombay, India.{\tt \footnotesize \{anantjoshi,maulik.bhatt,arpita.sinha\}@iitb.ac.in}}%
}

\begin{document}

\maketitle
\thispagestyle{empty}
\pagestyle{empty}
\begin{comment}
\begin{abstract}	
	This paper addresses the problem of exploring a region using the Hilbert's space-filling curve in the presence of obstacles. No prior knowledge of the region being explored is assumed. An online algorithm is proposed which can implement evasive strategies to avoid obstacles comprising a single or two blocked unit squares placed side by side and successfully explore the entire region. The strategies are specified by the change in the waypoint array which robot going to follow. The fractal nature of the Hilbert's space-filling curve has been exploited in proving the validity of the solution. 
\end{abstract}
\end{comment}

\begin{abstract}	
	This paper addresses the problem of exploring a region using Hilbert's space-filling curve in the presence of obstacles. No prior knowledge of the region being explored is assumed. An online algorithm is proposed which can implement evasive strategies to avoid up to two obstacles placed side by side and successfully explore the entire region. The strategies are specified changing the waypoint array followed by the robot, locally at the hole. The fractal nature of Hilbert's space-filling curve has been exploited in proving the validity of the solution. Extension of algorithm for bigger obstacles is briefly shown.
\end{abstract}
\vspace{-2mm}
\section{Introduction}

A space-filling curve is a one-dimensional curve which passes through every point of an N-dimensional region \cite{c5}. Rigorously, it is a continuous mapping from $[0,1]$ to an $n$-dimensional Euclidean space which has a positive $n$-dimensional Jordan content \cite{sagan}. 
%Peano was the first to discover a space-filling curve in 1890. 
In 1891, D. Hilbert proposed Hilbert's space-filling curve (hereby referred to as HC). %which gave insight into the geometry and construction of such curves.  

The HC has interesting properties which are exploited in various applications. Being a space-filling curve it covers all points in a two-dimensional square space. This intrinsic property of all space-filling curves attracts many applications where an area needs to be covered by an agent like tool path planning \cite{c6}, area coverage problems \cite{c7} and optimization problems where HC is used to convert a multi-dimensional problem to a one dimension \cite{c2}.   
It has a locality preserving property by virtue of which images of points that are close to each other, are also close to each other \cite{lpp,bader,lpp5}.
%\cite{lpp8} proposes a new space-filling curve which has even better locality preservation. 
Locality preservation is sought after in computer science applications involving parallel computing, data storage and indexing of meshes \cite{bader,lpp6,cluster,lpp7}.  
Its fractal nature allows us to zoom in and out of the curve and still obtain the same pattern. This inspires application to search problems where the space to be explored has patches which are more interesting or crucial than others \cite{c9}.  The fractal nature of the HC is exploited to explore patches of higher interest at a higher level of refinement resulting in higher resolution of observation. 
%\vspace{-10pt}

There is a myriad of literature addressing robotic search and path planning problems \cite{pathsurvey}. HC finds its application in this area too by virtue of its attractive properties which lead to novel and interesting solutions to conventional problems \cite{c5,c9}. However, both consider an environment without any obstacles for the robotic agent. 
Problems considering the robotic exploration of a space containing obstacles, along a space-filling curve are interesting, because it may not always be trivial to find a path around the obstacles and continue on the curve. 
\cite{sierpinski} presents a solution to such a problem using a Sierpinski curve but this method requires knowledge of the domain to be explored in advance. A solution which can be implemented online would be a welcome candidate for an autonomous agent.
%\vspace{-30pt}

\begin{figure}
	\centering
	\begin{subfigure}{.2\textwidth}
		\centering
		\includegraphics[scale=0.08, angle=270]{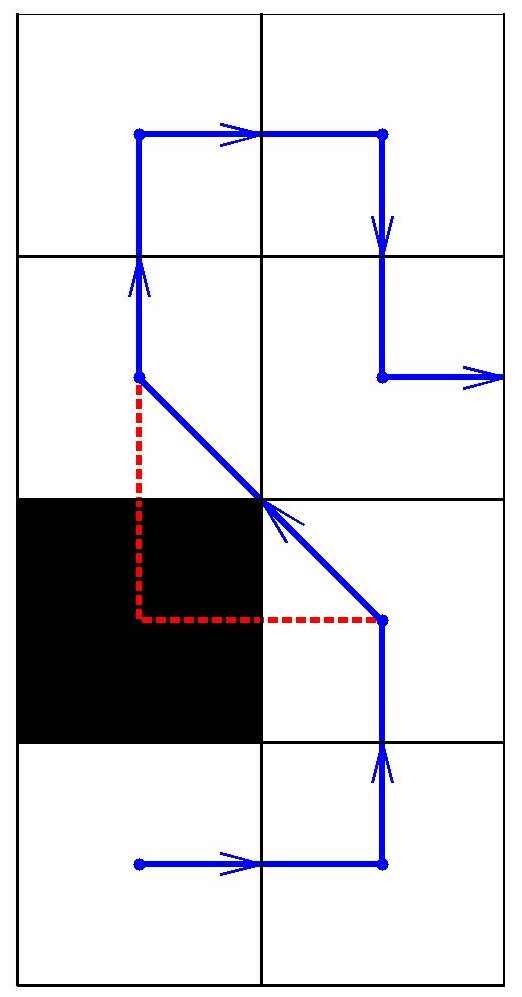}
		\caption{}
		\label{SNf1}
	\end{subfigure}
	\begin{subfigure}{.2\textwidth}
		\centering
		\includegraphics[scale=0.08, angle = 270]{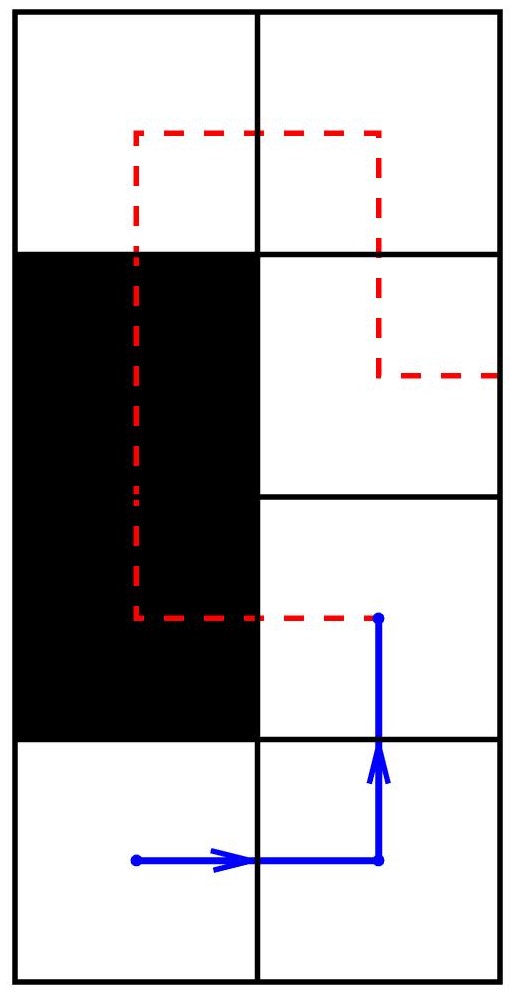}
		\caption{}
		\label{SNf2}
	\end{subfigure}
	\caption{(a) Strategy for single node({\cite{SN}}) (b)Strategy fails}
	\label{SNf}
\end{figure}

We consider the problem of robotic exploration, i.e., of finding a path for an autonomous agent to traverse, such that it is able to explore a given region fully. We employ the HC in this endeavour. Such an online strategy is presented in \cite{SN}, which divides the space into a grid based on the sensor resolution of the agent, and plans a path when the obstacle occupies one grid square. If there is a bigger obstacle, these strategies are unable to guarantee evasion (for eg. \cref{SNf}).
We extend this work by considering obstacles of bigger size and establishing online evasion strategies for the same. We rely on fractal nature of the HC and exploit the grammar \cite{bader} of the HC to design and verify our strategies. We redefine the methods presented in \cite{SN} since, with some modifications, they are able to evade holes with two blocked nodes.
We work under the assumption that two squares in the grid are blocked. The techniques presented in this work can be extended to other space-filling curves by using the grammar of those curves. We present simulation and hardware experiments to demonstrate our strategies.
%For the experimental validation and implementation of the algorithm, the strategy used was similar to that of the \cite{borkar2019aerial}.

\section{Preliminaries}

\subsection{Mathematical Preliminaries}
\begin{figure}
	\centering
	\begin{subfigure}{.15\textwidth}
		\centering
		\includegraphics[scale=0.065]{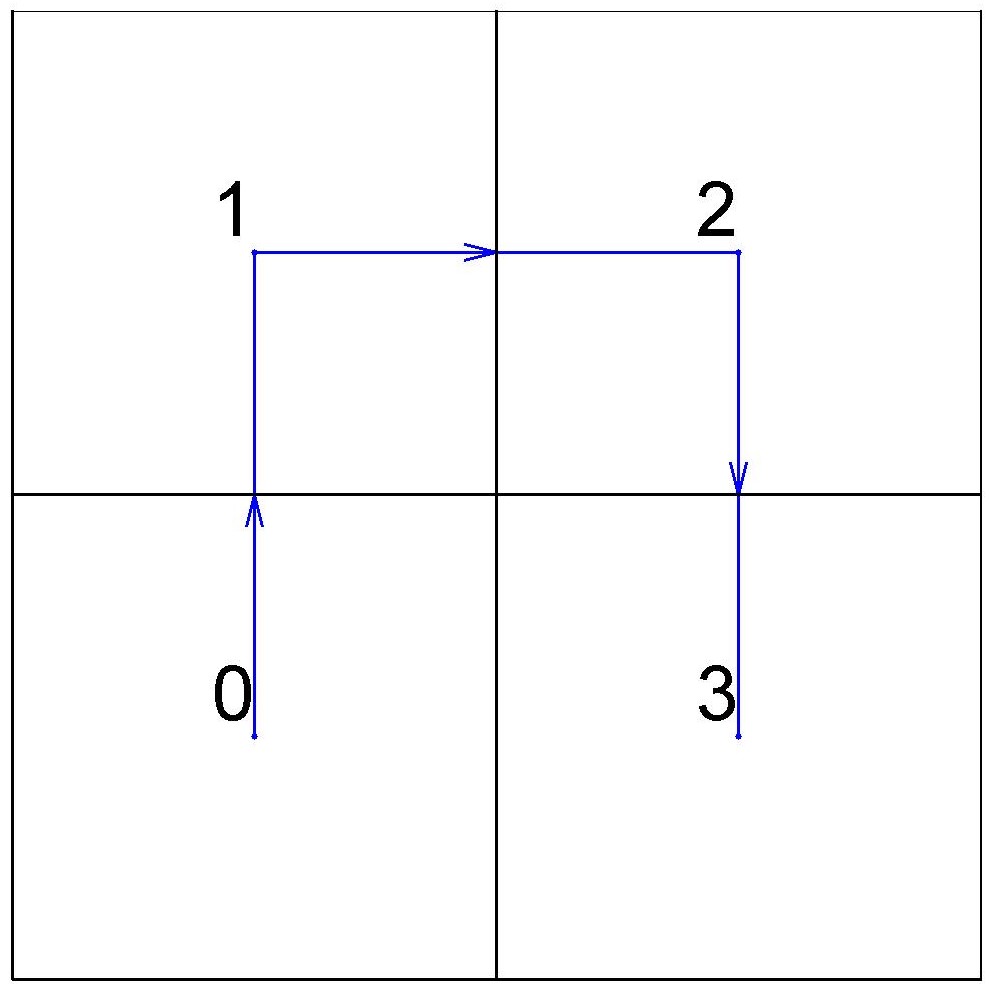}
		\caption{}
		\label{fh2}
	\end{subfigure}
	\begin{subfigure}{.15\textwidth}
		\centering
		\includegraphics[scale=0.065]{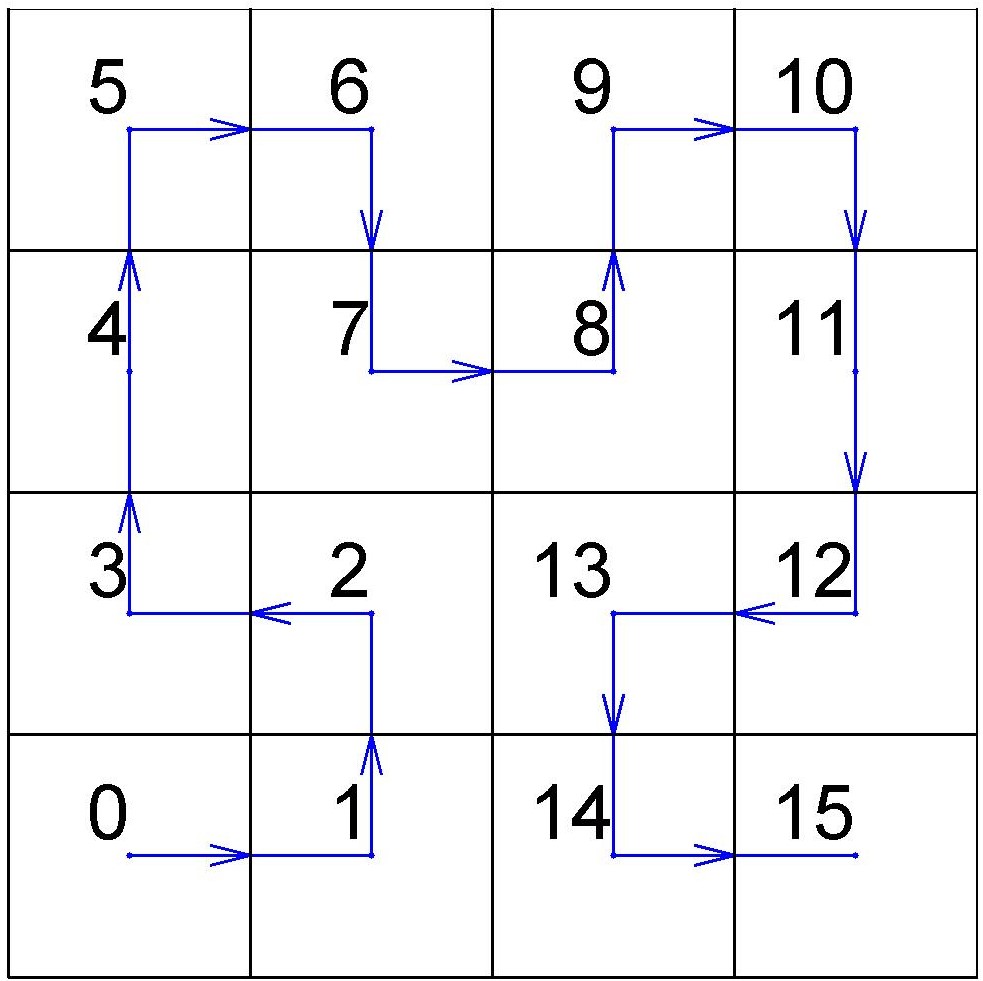}
		\caption{}
		\label{fh3}
	\end{subfigure}
	\caption{\small Approximated (a)first (b)second order HC}
	\label{fh}
\end{figure}

In this paper, we will be dealing with the HC which maps from the unit interval $\mathbf{I}=[0,1]$ to the unit square $\mathbf{Q}=[0,1]\times[0,1]$. 
%The curve is generated using a sequence of nested closed intervals. 
$\mathbf{I}$ is first divided into four equal sub-intervals which are mapped onto four congruent sub-squares of $\mathbf{Q}$ such that adjacent sub-intervals map to adjacent sub-squares. %For eg. the first partition is made as $[0,0.25] \leftarrow [0,0.5]\times[0,0.5], [0.25,0.5] \leftarrow [0,0.5]\times[0.5,1], [0.5,0.75] \leftarrow [0.5,1]\times[0.5,1], [0.75,1] \leftarrow [0.5,1]\times[0,0.5]$ . 
 Repeating this process on the sub-intervals and corresponding sub-squares for $N$ iterations produces a HC of order $N$. On repeating this procedure infinitely, the complete HC is generated \cite{sagan}. 
 We use an approximated version of the HC \cite{SN} (Fig. \ref{fh}). For an $N^{th}$ order HC, $4^N$ points in $\mathbf{I}$ namely $\lambda_i = \frac{i}{4^N}$ for $i = 0,1,2...4^N - 1$ are used.  
Under the approximation used, these points are mapped to the centre-points of the subsquares obtained when $\mathbf{Q}$ is divided into $4^N$ congruent sub-squares. Every node maps to a unique subsquare and likewise, every subsquare corresponds to a unique node. 
%An inverse mapping $\mathcal{N}^{-1}$ is defined in algorithmic form in Algorithm \ref{aninv} which maps back from the node locations in $\mathbf{Q}$ to the points in $\mathbf{I}$. Both the maps, $\mathcal{N}$ and $\mathcal{N}^{-1}$ are injective 
The images of these points in $\mathbf{Q}$ will be referred to as nodes of the $N^{th}$ order HC.
Any node of an $N^{th}$ order HC will be addressed by \textit{node number} defined to be the index of the point in $\mathbf{I}$ which maps to that subsquare. 
For instance, the node number of the subsquare to which $\lambda_2$ maps is 2, and this node will be addressed as `node 2'. 

\begin{figure}[]
	\centering
	\begin{subfigure}{.1\textwidth}
		\centering
		\includegraphics[scale=0.15]{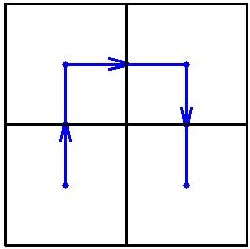}
		\caption{}
		\label{HilbertH}
	\end{subfigure}
	\begin{subfigure}{.1\textwidth}
		\centering
		\includegraphics[scale=0.15]{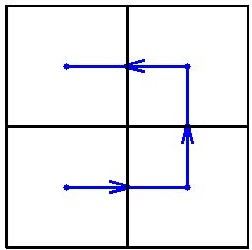}
		\caption{}
		\label{HilbertA}
	\end{subfigure}
	\begin{subfigure}{.1\textwidth}
		\centering
		\includegraphics[scale=0.15]{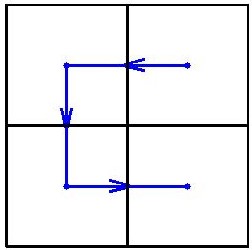}
		\caption{}
		\label{HilbertB}
	\end{subfigure}
	\begin{subfigure}{.1\textwidth}
		\centering
		\includegraphics[scale=0.15]{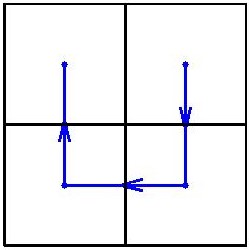}
		\caption{}
		\label{HilbertC}
	\end{subfigure}
	\caption{$\mathcal{H,A,B,C}$ orientations of a first order HC }
	\label{horientation}
\end{figure}

The HC being a fractal can have exactly four possible orientations denoted by $\mathcal{H,A,B,C}$ (fig. \ref{HilbertH}, \ref{HilbertA}, \ref{HilbertB}, \ref{HilbertC} respectively) \cite{bader}. A HC of one orientation can be transformed into another by rotation and reflection using the maps $L_0,L_1,L_2$ and $L_3$ given in \cref{tL}. When we refer to any HC, we refer to it's $H$ orientation by default, unless otherwise specified.
\small
\begin{table}[]
    \centering
    \begin{tabular}{c|c|c|c}
         $ L_0 := \begin{bmatrix}	0 & 1\\1 & 0 \end{bmatrix} $ & $L_1 := \begin{bmatrix}	1 & 0\\0 & 1 \end{bmatrix}$ & $L_2 := -L_0$ & $L_3 := -L_1$  \\ 
    \end{tabular}
    \caption{Maps defining HC Grammar}
    \label{tL}
\end{table}
\normalsize
\begin{comment}
\begin{minipage}{0.23\textwidth}
	\begin{flalign}
	& L_0 := \begin{bmatrix}	0 & 1\\1 & 0 \end{bmatrix} & \label{e1} \\	
	& L_1 := \begin{bmatrix}	1 & 0\\0 & 1 \end{bmatrix} & \label{e2} 
	\end{flalign}
\end{minipage}
\hfill
\begin{minipage}{0.23\textwidth}
	\begin{flalign}
	& L_2 := \begin{bmatrix}	0 & -1\\-1 & 0 \end{bmatrix} & \label{e3} \\
	& L_3 := \begin{bmatrix}	-1 & 0\\0 & -1 \end{bmatrix} & \label{e4} 	
	\end{flalign}
\end{minipage}
%\addtocounter{equation}{4}
\normalsize
\end{comment}
\begin{remark}
\label{r1}
\textit {$L_i^TL_i = \mathcal{I}$ hence $L_i$ represent either rotation or reflection for $i=1,2,3,4$ ($\mathcal{I}$ is the identity matrix).} 
\end{remark}
In \cref{fh} it can be seen that the second order curve is made up of 4 first order curves suitably transformed using \cref{tL}. This property along with a finite number of possible orientations for an HC (fig. \ref{horientation}) highlight the fractal nature of the HC. Due to this fractal nature any HC of order $N_1$ can similarly be constructed from $4^{N_1 - N_2}$ HC of order $N_2$ ($N_1 > N_2$). Thus, intuitively any higher order HC can be thought to be composed of lower order HC.

 Nodes can be classified as corner or non-corner nodes \cite{SN}. Corner nodes are nodes which enter or exit any first order HC. 
 Node 0 and node 3 are corner nodes in Fig. \ref{fh2}. 
One more classification is introduced, where 5 types of nodes are defined called $\tau_1,\ldots,\tau_5$. The procedure to check the type of the node is as follows: suppose $i$ is the node number of the node in concern. Define two vectors: one as a vector($\vv{v}_1$) from $(i-1)^{th}$ node to $i^{th}$ node and the other as a vector($\vv{v}_2$) from $i^{th}$ node to $(i+1)^{th}$ node. Define $r := \frac{\vv{v}_1\cdot\vv{v}_2}{|\vv{v}_1||\vv{v}_2|}$ to be used in the classification. $\%$ will be used for the mathematical operator $\mod$ (modulo).

\begin{table}[H]
	\centering
	\begin{tabular}{c || c || c || c || c}
		$\tau_1 : r < 1$ & $\tau_2 : r=0$ & $\tau_3 : r>0$ &
		$\tau_4 : r=1$ & $\tau_5 : r<0$ \\
	\end{tabular}
	
	\caption{Sets based on path passing through the node}
	\label{ttau}
\end{table}

\begin{comment}

{\centering
$\frac{\vv{v}_1\cdot\vv{v}_2}{|\vv{v}_1||\vv{v}_2|}<1$ $\Rightarrow$ i is a \textit{type 1} node \\ 
$\frac{\vv{v}_1\cdot\vv{v}_2}{|\vv{v}_1||\vv{v}_2|}=0$ $\Rightarrow$ i is a \textit{type 2} node\\
$\frac{\vv{v}_1\cdot\vv{v}_2}{|\vv{v}_1||\vv{v}_2|}>0$ $\Rightarrow$ i is a \textit{type 3} node\\ 
$\frac{\vv{v}_1\cdot\vv{v}_2}{|\vv{v}_1||\vv{v}_2|}=1$ $\Rightarrow$ i is a \textit{type 4} node\\
$\frac{\vv{v}_1\cdot\vv{v}_2}{|\vv{v}_1||\vv{v}_2|}<0$ $\Rightarrow$ i is a \textit{type 5} node\\}
\end{comment}

\subsection{Terminology Used}

\begin{comment}
\begin{itemize}
	\item Order of a HC(done), natural order(done), normal Hilbert curve
	\item Hole(done)
	\item Corner node
	\item Visited node
	\item Covered node(done)
	\item Node(done), blocked, unblocked, available (all done)
	\item Node number(done)
	\item Connected and disconnected second order curve
	\item Neighbours, pre-neighbour, post-neighbour
	\item Edge connected (done)
	\item Vertex connected (done)
	\item $n_2$ strictly greater than $n_1$ always!
	\item Hilbert curve (done)
	\item Hole is evaded (done)
	\item Evasion strategy (done)
	\item Path (done)
\end{itemize}
\end{comment}

The \textit{natural order} of the HC refers to the HC followed in ascending order of its nodes. 
%When the co-ordinates of a node($n$) are referred to, it means the co-ordinates of $\mathcal{N}(n,N)$ in $\mathbf{Q}$. 
The distance between any two nodes $n_1,n_2$ defined as $d(n_1,n_2)$ will be the Euclidean distance between them. 
The obstacle blocking the nodes is referred to as a \textit{hole}. 
%It will be referred to by the node numbers of the nodes it blocks. 
%These nodes that it blocks are referred to as \textit{blocked} nodes. 
%All the nodes that are not blocked are called \textit{available} nodes. 
%Once the agent reaches the co-ordinates of a node, the node will be said to have been \textit{visited} or \textit{covered} by the agent. 
Two nodes are said to be \textit{edge connected} if they share a common edge and \textit{vertex connected} if the sub-squares share a common vertex but not a common edge. 
The \textit{neighbours} of a node is the set of all those nodes which are edge connected or vertex connected to it. 
%A \textit{path} is a sequence of nodes such that all nodes can be visited successively one after the other that is, a sequence of nodes such that each node (except the first in the sequence) is a neighbour of the previous one. 
A hole is said to have been \textit{evaded} if all nodes of the HC except those blocked by the hole are covered. 
An \textit{evasion strategy} describes a path to successfully evade a hole. 
%\subsection{Classification of Nodes of the HC}
%We define a function $b(n)$ which returns the second last digit in the quaternary representation of the node.
\section{Main Result}

\subsection{Problem Formulation}

The problem of a robotic agent on an exploration task moving on the nodes of an $N^{th}$ order HC in $\mathbf{Q} = [0,1]\times[0,1]$ is considered, in the presence of obstacles. It is assumed that the sensing range of the agent is limited to the neighbouring nodes of the agent's current node. It is assumed that the obstacle covers two or fewer nodes of the HC. An online algorithm is presented to implements strategies that evade the blocked nodes. The evasion strategies make sure that all available nodes except the ones blocked by the obstacle are covered by the agent and that the agent exits $\mathbf{Q}$ at the last node of the HC.
It will be assumed that for a HC of order $N$, the first node, which is node 0 (also called entry node) and the last node, which is node $4^N-1$ (also called exit node) of the HC are available otherwise the agent cannot enter and exit the curve. 
\vspace{-1mm}
\subsection{Proposed Solution to Evade upto Two Blocked Nodes} 
Evasion strategies have been proposed to evade the holes under consideration.
When a hole is detected, a detour is planned that takes the agent around the hole and puts it back on the HC. The detour strategies are presented in \cref{tnb,t2,t3}. 
The strategies modify the order in which the HC is traversed.

\begin{figure}[]
	\centering
	\includegraphics[scale=0.4]{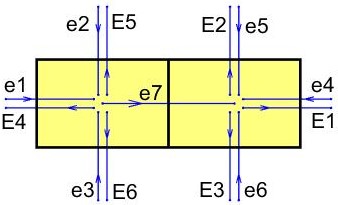}
	\caption{Possible ways to enter and exit a hole}
	\label{eE}
\end{figure}

Table \ref{tnb} contains strategies from \cite{SN} to evade a single blocked node, but suitably modified for our application. If there are two blocked nodes, they may be edge connected, vertex connected or not connected at all. For the latter two cases, evasion is done by strategies in \cref{tnb} (as we show further). For edge connected nodes, we begin by identifying how the hole may be entered or exited in a normal HC (\cref{eE} which shows blocked nodes in yellow and paths in blue). We see that the hole can be entered via one of 6 ways $e1$ through $e6$ and exited via $E1$ through $E6$. $e7$ is used only when both the blocked nodes are consecutive. This gives us a hint of how to devise evasion strategies for these cases. Some of these cases can be evaded by strategies of \ref{tnb} (for eg. $e2,e7,E2$) as we show further. However, it is clear that others cases (for eg. $e2,e7,E3$) require the development of some new strategies. Table \ref{t2} and \ref{t3} contain these new proposed strategies. Finally, an algorithm is presented to implement these strategies.

\subsection{Validity of Solution}

We will consider a HC of order $N$ in this subsection. It will be assumed that nodes $0$ and $4^N - 1$ are available, that is, if a node $n_b$ is assumed to be blocked, $n_b \notin \{0,4^N - 1\}$. 
\begin{comment}
We will define $d_0:=2^{-N}$ which is the distance between consecutive nodes. To begin with, for any two nodes $n_1,n_2$ and for any single node $n$, we will define the following for easy reference:
\begin{equation}
\label{dist1}
\begin{aligned}
    & \d_1:=d(n_1-1,n_1+2) & \d_2:=d(n_1-1,n_2)\\
    & \d_3 := d(n_1,n_2+2) & \d_4:=d(n_1-2,n_1) \\
    & \d_5:=d(n_2,n_2+2) &  \d_6:=d(n-1,n+1) \\
    & \d_7:=d(n-1,n+3) &  \d_8 := d(n-1,n-3) \\
\end{aligned}
\end{equation}
\end{comment}
We will first restate the main result of \cite{SN} and highlight some of its properties.

\vspace{-15pts}

\noindent \begin{lemma}
	\label{L3} 
	If there is a hole comprising a single blocked node($n_b$) in the HC, there always exists a path from node $n_b - 1$ to node $n_b + 1$. This path passes only though edge connected neighbours of the blocked node.
\end{lemma}

\noindent \textit{Proof:} 
	Any single blocked node $n_b$ can be evaded by strategies described in \cref{tnb}, which by virtue of their construction use only edge connected neighbours of the blocked node. $\blacksquare$
	
%We will now make certain observations regarding the evasion strategies in \cref{tnb}. 
%Consider a node $n_b \in S_7$. 
% It is a non-corner node, for which $n_b-1$ and $n_b+1$ are both edge connected to $n_b$ and vertex connected to each other, hence a direct path from node $n_b-1$ to $n_b+1$ can be taken to evade $n_b$. Hence evasion strategy for $n_b \in S_7$ (non corner node) requires only nodes $n_b - 1$ and $n_b + 1$ to be available. 
% In all other cases to avoid $n_b$, a detour is taken through either $n_b-3$ or $n_b+3$. 
%All evasion strategies in \cref{tnb} only require edge connected nodes. 
From the grammar of the HC, nodes $n_b+2$ and $n_b-2$ can never be edge connected to $n_b$. Thus, from \cref{L3} availability of nodes $n_b - 2$ and $n_b + 2$ is not required in any of the evasion strategies of \cref{tnb}. 
The evasion strategies do not use nodes $n \le n_b - 4$ or $n \ge n_b + 4$. The figures in \cref{tnb} provide an illustration.
\begin{comment}
\begin{lemma} The following statements regarding the evasion strategies in \cref{tnb} for a single blocked node $n_b$ are true:
    \label{L4} 
	\begin{enumerate}[(i)]
	 \item Evasion strategy for $n_b \in S_7$ (non corner node) only requires availability of nodes $n_b - 1$ and $n_b + 1$
	\item Availability of nodes $n_b - 2$ and $n_b + 2$ is not required in any of the evasion strategies.
	\item Availability of nodes($n$) such that $n \le n_b - 4$ or $n \ge n_b + 4$ is not required in any of the evasion strategies.
	\end{enumerate}
\end{lemma}

\noindent \textit{Proof:}
    A node $n_b \in S_7$ is a non-corner node, for which $n_b-1$ and $n_b+1$ are both edge connected to $n_b$ and vertex connected to each other, ($d(n_b-1,n_b+1) = \sqrt{2}d_0$) hence a direct evasion path from node $n_b-1$ to $n_b+1$ can be taken to evade the node, proving (i). All evasion strategies only require edge connected nodes. %in the same {\color{red}{same and also the ones immediately before and after this one? I believe the next sentence is true even without this sentence.} } first order HC to be available. 
    From the grammar of the HC, nodes $n_b+2$ and $n_b-2$ can never be edge connected to $n_b$, proving (ii).  Nodes($n$) $n \le n_b - 4$ or $n \ge n_b + 4$ lie in different first order HC from the one in which $n_b$ lies, proving (iii). \hfill $\blacksquare$

The illustrations in \cref{tnb} give an intuition of \cref{L4}.
\end{comment}
We proceed to the case when holes consist of two blocked nodes $n_1$ and $n_2$ ($n_1 < n_2$). To begin, we divide them into four sets (Table \ref{tdiff}) based on the difference in node number. 
\begin{table}[]
	\centering
	\begin{tabular}{c|c||c|c}
		$\mathfrak{A}$ & $n_2 = n_1 + 1$ & $\mathfrak{B}$   &$n_2 = n_1 + 2$ \\ \hline
		$\mathfrak{C}$ & $n_2 = n_1 + 3$ & $\mathfrak{D}$ & $n_2 \ge n_1 + 4$  \\
	\end{tabular}
	
	\caption{Sets based on node number difference}
	\label{tdiff}
\end{table}
Holes falling in $\mathfrak{B}$ and $\mathfrak{D}$ and a certain subset of those falling in $\mathfrak{C}$ can be evaded by the strategies in \cref{tnb}. %when used as prescribed in Algorithm \ref{a1}. Showing this will be the objective of the next 3 Lemmas. 

\begin{lemma}
	\label{L5}
	 Holes consisting of exactly two blocked nodes $n_1$ and $n_2$ in (i) $\mathfrak{B}$ or (ii) $\mathfrak{D}$ can be evaded by strategies described in \cref{tnb}. The nodes between the blocked nodes will also be covered. 
\end{lemma}
\noindent \textit{Proof:}
	We claim that for both cases $n_1$ and $n_2$ can individually be evaded by strategies in \cref{tnb}. We will prove this by showing that all nodes required for evasion are indeed available. $n_2 = n_1 +2 \Rightarrow n_1 - 3 < n_1 < n_2 < n_1 + 3 \Rightarrow n_1+3$ and $n_1-3$ are available in both cases. Thus a path from $n_1-1$ to $n_1+1$ exists evading $n_1$. Similarly, $n_2-3$ and $n_2+3$ both available ($n_2 - 3 < n_1 < n_2 < n_2 + 3$) in both cases so a path from $n_2-1$ to $n_2+1$ exists evading $n_1$. In case of $\mathfrak{B}$ the only node in between is $n_1 + 1$ which is covered by the evasion for $n_1$ and in case of $\mathfrak{D}$ all nodes in between are covered since the normal HC if followed in between. $\blacksquare$
\begin{comment}
\noindent \textit{Proof:}
	Since the hole consists of exactly nodes $n_1$ and $n_2$, nodes $n_1 - 1$, $n_1 + 1$, $n_2 - 1$ and $n_2 + 1$ are available for both (i) and (ii). For (i) observe that $n_1 + 1 = n_2 - 1$. 
	
	a) For (i) $n_1 + 1$ is the only node in between $n_1$ and $n_2$. Hence if $n_1 + 1$ is covered, all nodes between $n_1$ and $n_2$ will be covered as that . For (ii) if node $n_1 + 1$ is covered, all nodes till and including $n_2 - 1$ will be covered since there are no blocked nodes in the path connecting them.
	
	b) Consider the case if only $n_1$ was the blocked node. Then by Lemma \ref{L3} there exists a path from $n_1 - 1$ to $n_1 + 1$. From Lemma \ref{L4} this path does not require availability of node $n_1 +2$ or nodes($n$) such that $n \le n_1 - 4$ or $n \ge n_1 + 4$. Hence even though $n_2$ is blocked, this path will continue to be available. Similarly the path from $n_2-1$ to $n_2 + 1$ will continue to remain available even though $n_1$ is blocked. 
	
	From (a) and (b), for both (i),(ii), there exists a path from $n_1 - 1$ to $n_2 + 1$ which also covers all nodes in between. $\blacksquare$
\end{comment}
The upcoming result will help us exploit the grammar of the HC for further proofs. Recall that the orientation $O$ of a first order HC can take one of four values $\mathcal{H,A,B,C}$. 

\begin{lemma}
\label{L6-a}
Given two consecutive first order HC in any $N^{th}$ order HC with orientations $O_1,O_2 \in \{\mathcal{H,A,B,C}\}$, one of the maps in \cref{tL} transforms them into two consecutive first order HC curves with orientation $\mathcal{H},O_3$ or $O_3,\mathcal{H}$ with $O_3 \in \{\mathcal{H,A,B,C}\}$ preserving the following:
\begin{enumerate}[(i)]
\item The distance between any two nodes is invariant under the transformation
\item The sequence in which the nodes are traversed
\item A corner node and non-corner node will continue to remain a corner node and a non-corner node respectively
\end{enumerate}
Similarly, $H,O_3$ can be transformed into $O_1,O_2$ using the inverse transformation while preserving the same properties. 
\end{lemma}

\noindent \textit{Proof:}
     A HC of any orientation can be transformed into a HC of any other orientation using the maps in \cref{tL} \cite{bader}. We choose a map $L$ such that $L(O_1)=\mathcal{H}$. Then $O_3:=L(O_2) \in \{\mathcal{H,A,B,C}\}$. Or we choose $L$ such that $L(O_2)=\mathcal{H}$ then $O_3 = L(O_1)$. Since $L_i$ represents pure rotation or reflection for all $i$, which does not change distances between points, the distance between any two nodes is invariant under the transformation. Under any of the transformations, the order in which the nodes are traversed does not change since they are just rotations or reflections, hence nodes which enter or exit the first order HCs continue to do so thus corner nodes remain corner nodes and similarly for non-corner nodes. Similarly, $\mathcal{H},O_3$ can be transformed to $O_1,O_2$ using $L^T$ (inverse of $L$). \hfill $\blacksquare$
     
For $n_1,n_2 \in \mathfrak{A}$ quite a few cases can be evaded using the strategies from \cref{tnb}. In \cref{ta} we will sub-divide $\mathfrak{A}$ into subcases using \cref{eE}. $\mathfrak{a}_1,\mathfrak{a}_2, \mathfrak{a}_3, \mathfrak{a}_4$ can be evaded using \cref{tnb} and $\mathfrak{a}_5$ requires strategies from \cref{t3}. 

\begin{table}[h]
	\centering
	\begin{tabular}{c|c|c|c|c|c}
		\multicolumn{4}{c|}{$\mathfrak{a}_1 :(e_1,E_2),(e_1,E_3),(e_4,E_5),(e_4,E_6)$} & \multicolumn{2}{c}{$\mathfrak{a}_3 :(e_3,E_3),(e_2,E_2)$} \\ \hline	\multicolumn{4}{c|}{$\mathfrak{a}_2 :(e_2,E_1),(e_3,E_1),(e_5,E_4),(e_6,E_4)$}
		& \multicolumn{2}{c}{$\mathfrak{a}_4 :(e_1,E_1),(e_4,E_4)$}  \\ \hline 	\multicolumn{6}{c}{$\mathfrak{a}_5 :(e_2,E_3),(e_3,E_2),(e_5,E_6),(e_6,E_5)$}\\
	\end{tabular}
	\caption{Subcases of $\mathfrak{A}$}
	\label{ta}
\end{table}

\begin{figure}
    \centering
    \begin{minipage}[b]{.3\linewidth}
       \subcaptionbox{$\mathcal{HH}$}{\includegraphics[scale=0.15,keepaspectratio]{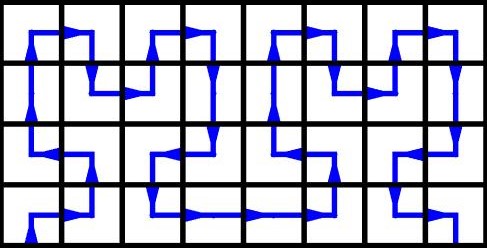}}
       \subcaptionbox{$\mathcal{HA}$}{\includegraphics[scale=0.15,keepaspectratio]{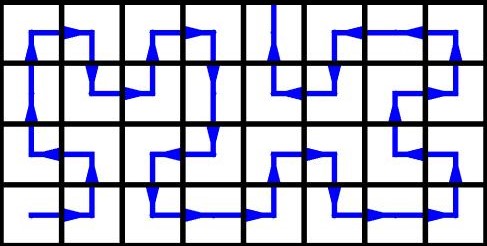}}
    \end{minipage}
    \begin{minipage}[t]{.18\linewidth}
        \subcaptionbox{$\mathcal{HB}$}{\includegraphics[scale=0.18]{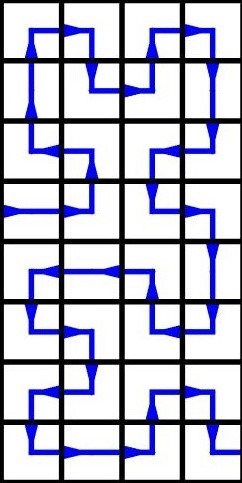}}
    \end{minipage}
    \begin{minipage}[t]{.18\linewidth}
        \subcaptionbox{$\mathcal{HC}$}{\includegraphics[scale=0.18]{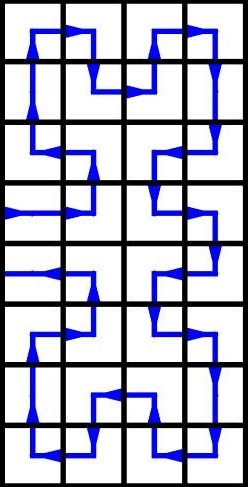}}
    \end{minipage}
    \caption{Combinations of two connected second order HC}
    \label{fig:secordersidebyside}
\end{figure}

\begin{lemma}
(i) Holes in $\mathfrak{a}_5$ can be evaded using \cref{t3}

\noindent (ii) Holes in $\mathfrak{a}_1,\mathfrak{a}_2,\mathfrak{a}_3,\mathfrak{a}_4$ can be evaded using \cref{tnb}
\label{L8-b}
\end{lemma}
\noindent \textit{Proof:}
    (i) First we will claim that both $n_1$ and $n_2$ belong to the same second order HC. Assume they don't, which means either $n_2 \% 16=0$ or $n_1 \% 16 = 15$. Assume $n_1 \% 16 = 15$. Now we will need to check all possible cases of two neighbouring second order HC to confirm this. However, due to its grammar \cite{bader} we need only check finite number of cases. From lemma \ref{L6-a} we need only check the four cases when the last node of a $\mathcal{H}$ orientation second order HC is blocked, and the next second order curve is of $\mathcal{H,A,B,C}$ orientation. Fig. \ref{fig:secordersidebyside} provides an illustration. It can be observed that in none of them, is the hole created of $\mathfrak{a}_5$ type. Checking similarly for $n_2 \% 16 = 0$ shows the claim. 
    
    We will prove the main result by claiming that all nodes required for evasion are available. We will show the availability of nodes in questions using HC grammar, \cref{L6-a} and \cref{t3}. 
    It can be seen from \cref{t3} that $n_1 \% 16 \in \{ 0,2,6,8,12,14\}$. For $n_1 \% 16 \in  \{2,6,8,12\}$ the proof is immediate by looking at a single second-order HC since all required for evasion are in the same-second order HC and available (refer \cref{t3}). For $n_1 \% 16 = 14$, we will look at all possible combinations of 2 connected second-order HC. We will assume that the hole blocks the last 2 nodes of the first second-order curve and check whether the nodes required for evasion are available. 
    However, we again need only check cases $\mathcal{HH}$, $\mathcal{AH,BH,CH}$ since all others can be converted to these using \cref{L6-a}. The case in question occurs only in the $\mathcal{AH}$ or $\mathcal{BH}$ case and the nodes required for evasion are available.  This can be proven on the exact same lines for $n_1\% 16 = 0$ hence result is proven for $\mathfrak{a}_5$.
    
    \noindent (ii) %We will prove this by claiming that all nodes required for evasion are available. We will show availability of nodes in question using \cref{eE}, HC grammar and \cref{L6-a}. 
    Suppose we would like to show the result for $\mathfrak{a}_1$. For two nodes $n_1$ and $n_2$ in $\mathfrak{A}$, either they both are in the same first order HC or in neighbouring first order HC, with $n_1$ being node 3 in the first one and $n_2$ being node 0 in the second one. 
    If $n_1$ and $n_2$ are in the same first-order HC, then they must satisfy $n_1 \% 4=0, n_2 \% 4 = 1$. In this case the nodes required for evasion are available. 
    To analyse the case when they're in neighbouring first order curves, similar to (i) check cases of neighbouring first order HC of orientations $\mathcal{HH,HA,HB,HC}$. It is seen that the nodes required for evasion are available. Hence result is proven for $\mathfrak{a}_1$ and can be proven similarly for the rest. \hfill $\blacksquare$ 

We will now establish a few facts (to be used in the content that follows) about two blocked nodes $n_1$ and $n_2$ in $\mathfrak{C}$. We will assume $N \ge 2$ since if $N=1$ (i.e. first order HC) then $n_1 = 0$ and $n_3=3$ which is not allowed since by assumption they are available. An $N^{th}$ order HC comprises $2^{2N - 2}$ first order HC. Let $n_1$ lie in the $N_F^{th}$ first order curve.
We will assume $N_F < 2^{2N - 2}$ in \cref{L6-b} since if $N_F = 2^{2N - 2}$ then $n_2 = 4^N - 1$ which must be available, by assumption. 
\begin{lemma}
\label{L6-b}
Consider two nodes $n_1$ and $n_2$ in $\mathfrak{C}$ as follows: assume $N \ge 2$ and let $n_1$ lie in the $N_F^{th}$ first order curve with $N_F \ne 2^{2N - 2}$. $n_1$mod4 $\in \{0,1,2,3\} $ and corresponding to each value:
    \begin{enumerate}[(i)]
    	\item If $n_1\%4$ $=0$ then $n_2$ lies in $N_F^{th}$ first order curve itself. They both are corner nodes and edge connected. 
    	\item If $n_1\%4$ $=1$ then $n_2$ lies in $(N_F + 1)^{th}$ first order curve. $n_1$ is a non-corner node and $n_2$ is a corner node and they are not edge connected.
    	\item If $n_1\%4$ $=2$ then $n_2$ lies in $(N_F + 1)^{th}$ first order curve. They both are non-corner nodes.
    	\item If $n_1\%4$ $=3$ then $n_2$ lies in $(N_F + 1)^{th}$ first order curve. $n_1$ is a corner node and $n_2$ is a non-corner node and they are not edge connected. 
    \end{enumerate}
\end{lemma}

\noindent \textit{Proof:}
    (i) Observe that $n_1$ and $n_2$ are the entry and exit nodes of a first order HC respectively, hence the result is established directly from the structure of a first order HC. We will use the following in further proofs:
    \begin{enumerate}
        \item If a node $n$ lies in the $N_F^{th}$ first order curve and $n$\%4=$k$ then $n = k + 4(N_F - 1)$
        \item For two nodes to be edge connected, the distance between them should be precisely equal to $d_0$ units. 
    \end{enumerate}
    \par (ii) $n_2 = 3 + 1 + 4(N_F - 1) = 4N_F + 0 $ hence $n_2$ lies in $(N_F + 1)^{th}$ first order curve. Assume that the $N_F^{th}$ first order HC is of $\mathcal{H}$ orientation. The distances between $n_1$ and $n_2$ is $d_0$ for all orientations $\mathcal{H,A,B,C}$ of the $(N_F + 1)^{th}$ curve. If $N_F^{th}$ HC is of a different orientation than $\mathcal{H}$, transform it into $\mathcal{H}$ orientation (\cref{L6-a}) and this transformation does not change distances between nodes. Hence $n_1$ and $n_2$ are not edge connected. $n_1$mod4=1,$n_2$mod4 $=0$ gives that $n_1$ is a non-corner node and $n_2$ is a corner node.
    Proof for (iii), (iv) follows similarly.
    %(iii) $n_2 = 3 + 2 + 4(N_F - 1) = 4N_F + 1 $ hence $n_2$ lies in $(N_F + 1)^{th}$ first order curve. $n_1$mod4=2,$n_2$mod4 $=1$ gives that both $n_1,n_2$ are non-corner nodes. 
    \hfill $\blacksquare$

We will group items (ii),(iii) and (iv) in the aforementioned list in \cref{L6-b} into $\mathfrak{c}_2$ (nodes such that both of them are non-corner nodes or exactly one of them is a corner node) and (i) into $\mathfrak{c}_1$ (both corner nodes). 

\begin{lemma}
	\label{L6}
	Holes in $\mathfrak{c}_2$ can be evaded using \cref{tnb}. 
\end{lemma}
 
\noindent \textit{Proof:}
	Since $n_1$ and $n_2 = n_1 + 3$ are the only blocked nodes,$n_1 - 1$, $n_1 + 1$, $n_2 - 1$ and $n_2 + 1$ are available. 
	
	\noindent a) If node $n_1 + 1$ is visited, $n_2 - 1$ will be visited since they are neighbours
	
	\noindent b) As shown earlier there will exist paths from $n_1 - 1$ to $n_1 + 1$ and $n_2 - 1$ to $n_2 + 1$ respectively if both are non-corner nodes.
	
	\noindent c) If exactly one of them is a corner node, %Prove that they don't share an edge by drawing figures and taking cases for $n_1$. Since they don't share an edge, a path according to (ref. Table II) from $n_1 - 1$ to $n_1 + 1$ and from $n_2 - 1$ to $n_2 + 1$ will always exist from (ref. Lemma 4). Node $n_1 + 2 = n_2 - 1$ is available by assumption. The path from node $n_1 - 1$ to $n_2 - 1$ to $n_2 + 1$ will be followed as prescribed by (ref. Algorithm I). 
	consider the case if only $n_1$ was the blocked node. Then from lemma \ref{L3} there exists a path from $n_1 - 1$ to $n_1 + 1$ which passes only through edge connected neighbours of $n_1$. Even though $n_2$ is blocked, this path will continue to be available since $n_1$ and $n_2$ are not edge connected neighbours (lemma \ref{L6-b}). Similarly for $n_2$ a path from $n_2 - 1$ to $n_2+1$ will exist and be available even though $n_1$ is blocked. 
	
	From (a), (b), (c), there exists a path from $n_1 - 1$ to $n_2 + 1$ which also covers all nodes in between. \hfill $\blacksquare$
	
Similarly like $\mathfrak{A}$, there exist 2 ways ($e_i$) to enter and 2 ways ($E_j$) to exit holes in $\mathfrak{c}_1$. In the following lemma, we will employ a similar approach as \cref{L8-b}. 

\begin{lemma}
\label{L8-c}
    Holes in $\mathfrak{c}_1$ can be evaded by using \cref{t2}. 
\end{lemma}

\noindent \textit{Proof (Sketch):}
   From lemma \ref{L6-b} the hole lies entirely in a single first order HC. We will look at all possible cases when such a hole occurs in a second order HC. For $n_1 \% 16 \in \{4,8\}$ all nodes required for evasion are observed to be available. For $n_1 \% 16 \in \{0,12\}$ analysing neighbouring second order HC, similar to lemma \ref{L8-b}, shows the result.  \hfill $\blacksquare$
\begin{thm}
    \label{th1}
    A HC having a hole consisting of at most two blocked nodes can be covered using strategies in \cref{tnb,t2,t3}.
\end{thm}
\noindent \textit{Proof:} Follows from \cref{L3,L5,L8-b,L6,L8-c}. \hfill $\blacksquare$

Theorem \ref{th1} can be easily extended to the case when there are multiple holes, but each hole consists of either a single blocked node, two edge connected blocked nodes or two vertex connected blocked nodes, and that the holes are not edge or vertex connected to each other (\cref{fig:expt}). 
\vspace{-2mm}
\section{Implementation}
\subsection{Philosophy and Salient Points of the Algorithm}

\begin{comment}
\begin{figure}[H]
	\centering
	\includegraphics[scale=0.13]{fig/multilpleevasion.jpg}
	\caption{Implementation of Evasion Strategies}
	\label{multipleblocked}
\end{figure}
\end{comment}

\begin{figure}[]
\begin{subfigure}{0.23\textwidth}
	\centering
	\includegraphics[width = 0.75\textwidth]{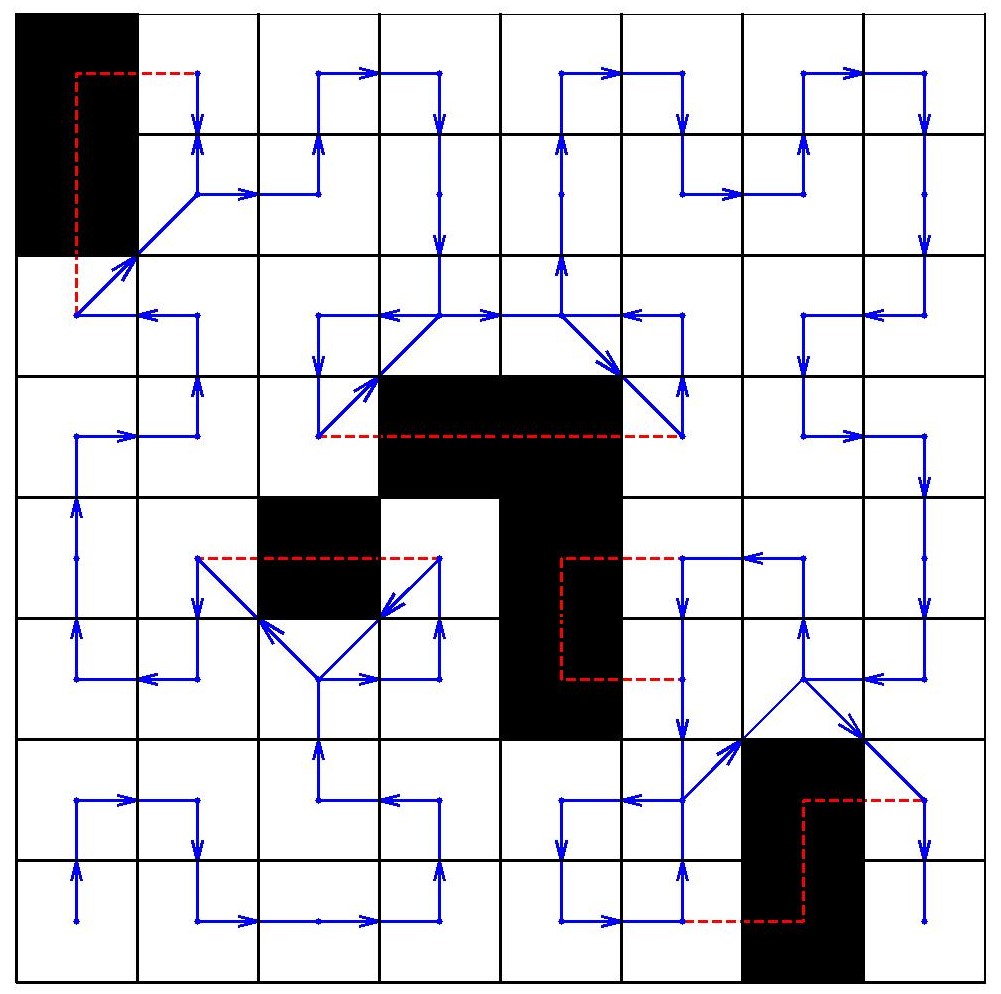}
	\caption{Computer Simulation}
	\label{fig:compsim}
\end{subfigure}
\begin{subfigure}{0.2\textwidth}
	\centering
	\includegraphics[width = 0.85\textwidth]{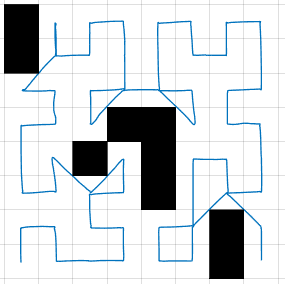}
	\caption{Hardware Experiment}
	\label{fig:hardexpt}
\end{subfigure}
\caption{Implementation of Evasion Strategies}
\label{fig:expt}
\end{figure}

\vspace{-1mm}
We present the \cref{alg:HC} to implement the strategies in \cref{tnb,t2,t3}. It takes as an input the order of the HC to be traveled and gives output an array containing coordinates of all nodes from node zero to the last node. Agent travels in a straight line between any two consecutive nodes. 
% This is an online algorithm and it requires no apriori knowledge of the blocked nodes. 
It checks if the next node is available or blocked, and commands the agent to either follow a certain strategy or continue on the HC. 
The node classification (as in \cref{ttau}) and algorithm design is done so as to evade maximum possible holes using strategies for a single blocked node (\cref{tnb}). For the cases when it is not possible, new strategies are introduced in \cref{t2,t3}. 
One of the examples where one obstacle strategy works is $(e_1,e_7,E_3)$, which can be also seen in the top left corner of the fig. 6a. In that case, the agent detects the obstacle at $n_1^{th}$ node and uses the strategy from \cref{tnb} to reach $(n_1+3)^{th}$ node. Now, the next node in $\mathcal{N}$ is $n_2$, it will be a $\tau_1$ node and from our algorithm, it will be skipped and the robot will go to $(n_2+1)^{th}$ node directly. 
In all the tables $\mathcal{N}$ represents the array containing all the nodes of HC which agent has to cover.
In \cref{tnb} the blocked node is either skipped or replaced by another available node to evade the hole.
In \cref{t2,t3} more then one consecutive elements of $\mathcal{N}$ is replaced by other available nodes to evade the hole. Note that the node classification in \cref{ttau} is not mutually exclusive as it is necessary by the virtue of the design of the algorithm.
In \cref{t2} it might be possible that the number of nodes to be added are more than the number of nodes to be replaced. In such cases. nodes which are already covered are replaced and the value of $n$ is decreased in order to preserve the total number of elements in $\mathcal{N}$. 
% {\color{blue}
% At times it may be the case that there are two blocked nodes, but only one of them is detected and the other lies undetected (for instance case B$_1$ in \cref{t2}) since we assume that the sensing range of the agent is restricted to the neighbours of the node it currently occupies. 
% In such a case, it is natural for an online algorithm to select a strategy for a single blocked node (\cref{tnb}). This however will lead to a problem since the second blocked node is required in the evasion strategy. The algorithm detects that the evasion strategy has been interrupted by the second blocked node and switches to a strategy from \cref{t1,t2}. The strategies have been designed by keeping this situation in mind thus the first displacement required by the new strategy (from \cref{t1,t2}) selected will always be the same as the first displacement required by the old strategy (from \cref{tnb}). 
% As a consequence, the first displacement of the new strategy will simply be skipped and the obstacle avoided.
% }
Computer simulation of evasion strategies is shown in \cref{fig:compsim}, where multiple holes have been considered in a HC of order 3. 

\vspace{-3mm}

\subsection{Experimental Validation}
\vspace{-1mm}
The implementation of the proposed algorithm was done in a similar manner to that of \cite{borkar2019aerial}. The implementation was done on a ground differential drive robot \textit{Firebird \RomanNumeralCaps{5}}. %\footnote{http://www.nex-robotics.com/products/fire-bird-xii/fire-bird-vatmega2560-robotic-research-platform.html}% 
The experiments were conducted in a motion capture environment using 8 \textit{VICON Vantage V5} cameras. %\footnote{https://www.vicon.com/products/camera-systems/vantage}. 
%The reflective markers, which are detected by the cameras to calculate position of the robot, were placed on the top of robot such that the centroid of the marker coincides with center of the robot. 
The data from \textit{VICON} cameras goes into the HP$\textsuperscript{\textregistered}$ Z-440 workstation. 
The Vicon Tracker Version 3.4 software
%\footnote{https://docs.vicon.com/display/Tracker34/Vicon+Tracker+User+Guide} 
in workstation extracts the position of the robot and  broadcasts it into the local Wi-fi. To access this data we use \textit{vicon-bridge}
%\footnote{http://wiki.ros.org/vicon bridge} 
node in ROS. It makes the data available in pose feedback form.  The controller for the robot is run as nodes in a ROS Indigo Igloo %\footnote{http://wiki.ros.org/indigo}
environment in Ubuntu 14.04 on a Lenovo$\textsuperscript{\textregistered}$Z51-70 Laptop (Intel$\textsuperscript{\textregistered}$ Core i7-5500U CPU, 2.4GHz and 16 GB RAM). Communication between laptop and  \textit{Firebird \RomanNumeralCaps{5}} is done using Xbee$\textsuperscript{\textregistered}$ modules. In order to avoid further complexity in detecting the obstacles, the location of the obstacles were known to the set up apriori but it was made available to the node which runs the algorithm only when the agent reaches in the neighbourhood of the blocked node. %$\textsuperscript{\textregistered}$ Pro Series 1 module %\footnote{https://www.digi.com/products/models/xbp24-awi-001j}

We model the robot hardware using unicycle kinematics $ \dot{x} = v(cos(\psi)),\dot{y} = v(sin(\psi)),\dot{\psi} = \omega $. We get the current state as feedback from \textit{vicon\_bridge} node in ROS. Let the desired state be $(x_d,y_d,\psi_d)$. The desired state is computed in \textit{bot\_controller} node equipped with algorithm \ref{alg:HC} (written in Python) running in ROS. Then, real-time input of linear velocity$(v)$ and angular velocity$(\omega)$ to the robot is given as following:
$ v = k_1\sqrt{(x_d -x)^2 + (y_d-y)^2} $ and
$ \psi_d = \tan^{-1}\left(\frac{y_d-y}{x_d-x}\right), \omega = k_2(\psi_d-\psi) $.
The values of $k_1$ and $k_2$ are 1 and 1.8 respectively.
We performed two experiments \cite{expt1,expt2} (web-links). Results of \cite{expt2} are portrayed in \cref{fig:hardexpt}. 

\begin{algorithm}[H]
    \caption{HC Modification to evade obstacles}
    \begin{algorithmic}[1]
    %\State Include:$\mathcal{N}$ (Equation \ref{enodefunc}), $\mathcal{N}^{-1}$ (Algorithm \ref{aninv}) 
		\State Input : $N = \ $order of HC
		\State Output : Array $\mathcal{N}$ of nodes from $\mathcal{N}(0)$ to $\mathcal{N}(4^N-1)$
		\State Initialize : $n=0$ 
			\While{$n \le 4^N-1$}
			\State Initialize $n_1 = n$
				\If{$n_1$ is blocked}
				    \If{$n_1 \in \tau_1$ }
				    \State Initialize $n_2 = n_1+1$
				        \If{$n_2$ is blocked and $n_2 \in \tau_2$}
				            \If{The combination, $(n_1,n_2)$, can be classified according to \cref{t3}}
				                \State Use the strategy described in \cref{t3}
				                \State Break
				            \Else
				                \State {$n \leftarrow n+1$}
				                \State Break
				            \EndIf
				        \ElsIf{$n_1 \in \tau_3$ }
				            \State Use the strategy described in \cref{tnb}
				            \State Break
				        \Else
				            \State {$n \leftarrow n+1$}
				            \State Break
				        \EndIf
				    \ElsIf{$n_1 \in \tau_4$}
				        \If{$n_1$\%2=0}
				            \State Initialize $n_2 = n_1+3$
				            \If{$n_2$ is blocked}
				                \State Use the strategy described in \cref{t2}
				                \State Break
				            \Else
				                \State Use the strategy described in \cref{tnb}
				                \State Break
				            \EndIf
				        \ElsIf{$n_1$\%2=1}
				            \State Initialize $n_2 = n_1-3$
				            \If{$n_2$ is blocked}
				                \State Initialize $n_2 = n_1$
				                \State $n_1 \leftarrow n_1-3$
				                \State Use the strategy described in \cref{t2}
				                \State Break
				            \Else
				                \State Use the strategy described in \cref{tnb}
				                \State Break
				            \EndIf
				        \EndIf
				    \Else
				        \State Pass
				    \EndIf
				\Else
				    \State Pass
			    \EndIf
			\State  {Visit $\mathcal{N}(n)$}
			\State $n \leftarrow n+1$
			\EndWhile
    \end{algorithmic}
    \label{alg:HC}
\end{algorithm}

\section{Conclusions}

The problem of exploring a region with holes using the HC has been considered. The holes are assumed to covers two or fewer nodes of the HC and an online algorithm is presented which implements strategies that evade any such obstacle. To prove the evasion of nodes, the fractal nature of the HC has been exploited. Computer simulations and hardware experiments demonstrate tractability of proposed algorithm. Extension of algorithm with more blocked nodes is shown through numerical simulation and hardware experiments. 

\begin{table}[H]
	\centering
	\begin{tabular}{|m{1.4cm}|m{2.5cm}|m{0.7in}|} \hline \hline		
		Node Type & Change in Path & Figure \\ 
		\hline \hline %\vspace{10em}

		%----

		\begin{tabular}{@{}c@{}} 
			$n_b \in \tau_2 \cup \tau_5$ \\
		\end{tabular}  
		&
		\begin{tabular}{@{}c@{}}  $n \leftarrow n-1$  \end{tabular}
		& 
		\parbox[c]{1.5in}{
			\includegraphics[scale=0.06, keepaspectratio, angle = 270]{fig/SN2.jpg}		
		} 
		\\ \hline

		\begin{tabular}{@{}c@{}} 
		$n_b \in \tau_3$ \\ $n_b\%2 = 1$
		\end{tabular} 
		&
		\begin{tabular}{@{}c@{}} $ \mathcal{N}(n_b) \leftarrow \mathcal{N}(n_b-3) $ \end{tabular}
		& 
		\parbox[c]{1.5in}{
			\includegraphics[scale=0.09, keepaspectratio]{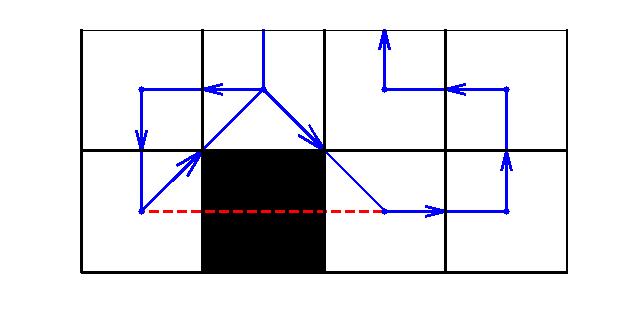}		
		} \\ 
		
		\hline

		\begin{tabular}{@{}c@{}} 
		$n_b \in \tau_3$ \\ $n_b\%2 = 0$
		\end{tabular} 
		&
		\begin{tabular}{@{}c@{}} $ \mathcal{N}(n_b) \leftarrow \mathcal{N}(n_b+3) $\end{tabular}
		& 
		\parbox[c]{1.5in}{
			\includegraphics[scale=0.09, keepaspectratio]{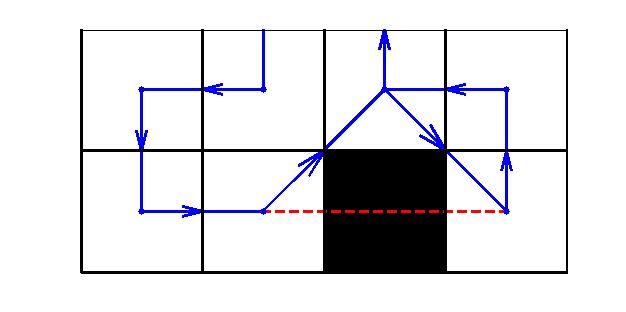}		
		} \\ 
		
		\hline
		%----
	
	\end{tabular}
	\caption{Evasion Strategies for Single Blocked Node ($n_b$) \footnotesize{$n$ is as used in algorithm \ref{alg:HC}}}
	\label{tnb}
\end{table}

\begin{comment}
\begin{table}[H]
	\centering
	\begin{tabular}{|m{0.45cm}|m{1.4cm}|m{2.5cm}|m{0.7in}|} \hline \hline		
		Name  & Node Type & Change in Path & Figure \\ 
		\hline \hline %\vspace{10em}

		%----
		
		$\mathscr{S}_1$ &		
		\begin{tabular}{@{}c@{}} 
			$n_b \in \tau_2 \cup \tau_5$ \\
		\end{tabular}  
		&
		\begin{tabular}{@{}c@{}}  $n \leftarrow n-1$  \end{tabular}
		& 
		\parbox[c]{1.5in}{
			\includegraphics[scale=0.06, keepaspectratio, angle = 270]{fig/SN2.jpg}		
		} 
		\\ \hline

		$\mathscr{S}_2$ &		
		\begin{tabular}{@{}c@{}} 
		$n_b \in \tau_3$ \\ $n_b\%2 = 1$
		\end{tabular} 
		&
		\begin{tabular}{@{}c@{}} $ \mathcal{N}(n_b) \leftarrow \mathcal{N}(n_b-3) $ \end{tabular}
		& 
		\parbox[c]{1.5in}{
			\includegraphics[scale=0.09, keepaspectratio]{fig/SN4.jpg}		
		} \\ 
		
		\hline
		
		$\mathscr{S}_3$ &		
		\begin{tabular}{@{}c@{}} 
		$n_b \in \tau_3$ \\ $n_b\%2 = 0$
		\end{tabular} 
		&
		\begin{tabular}{@{}c@{}} $ \mathcal{N}(n_b) \leftarrow \mathcal{N}(n_b+3) $\end{tabular}
		& 
		\parbox[c]{1.5in}{
			\includegraphics[scale=0.09, keepaspectratio]{fig/SN6.jpg}		
		} \\ 
		
		\hline
		%----
	
	\end{tabular}
	\caption{Evasion Strategies for Single Blocked Node ($n_b$) \footnotesize{$n$ is as used in algorithm \ref{alg:HC}}}
	\label{tnb}
\end{table}
\end{comment}

\begin{table}[H]
	\centering
	\begin{tabular}{|m{1.4cm}|m{2.8cm}|m{1in}|} \hline \hline		
		Node Type & Change in Path & Figure \\ 
		\hline \hline %\vspace{10em}

		%----

		\begin{tabular}{@{}c@{}} 
			$n_1 \in \tau_4$ \\ $n_2 \in \tau_2 $ \\ $n_1\%2 = 0$ \\ $n_2 = n_1+3$
		\end{tabular}  
		&
		\begin{tabular}{@{}c@{}}   $d_1:=d(n_1+4,n_1-2)$ \\ $d_2:=d(n_1+4,n_1-4)$ \\ if($d_1<d_2$) \\ $ \mathcal{N}(n_1-1,...,n_1+3) $\\ $\leftarrow$ $\mathcal{N}(n_1-2, n_1+4,$ \\ $n_1+2, n_1+1, n_1+2)$ \\$n \leftarrow n-1$ \\ elif($d1>d2$)  \\ $ \mathcal{N}(n_1-1,...,n_1+3) $\\ $\leftarrow$ $\mathcal{N}(n_1-4, n_1+4,$ \\ $n_1+2, n_1+1, n_1+2)$ \\$n \leftarrow n-1$  \end{tabular}
		& 
		\parbox[c]{1.5in}{
			\includegraphics[scale=0.12, keepaspectratio]{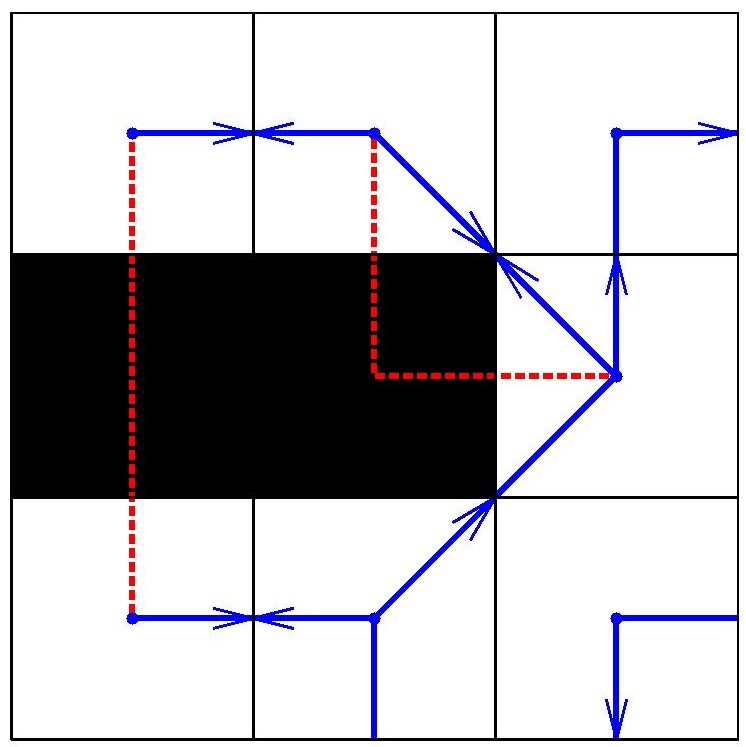}		
		} 
		\\ \hline

		\begin{tabular}{@{}c@{}} 
		$n_1 \in \tau_2 $ \\ $n_2 \in \tau_4$ \\ $n_2\%2 = 1$ \\ $n_2 = n_1+3$
		\end{tabular} 
		&
		\begin{tabular}{@{}c@{}}  $d_1:=d(n_2-4,n_2+2)$ \\ $d_2:=d(n_2-4,n_2+4)$ \\ if($d_1<d_2$) \\ $\mathcal{N}(n_2-2,...,n_2)$ \\ $\leftarrow$ $\mathcal{N} (n_2-2, n_2-4, $\\ $n_2+2$) \\$n \leftarrow n-2$ \\  elif($d1>d2$) \\ $\mathcal{N}(n_2-2,...,n_2)$ \\ $\leftarrow$ $\mathcal{N} (n_2-2, n_2-4, $\\ $n_2+4$) \\ $n \leftarrow n-2$\end{tabular}
		& 
		\parbox[c]{1.5in}{
			\includegraphics[scale=0.12, keepaspectratio]{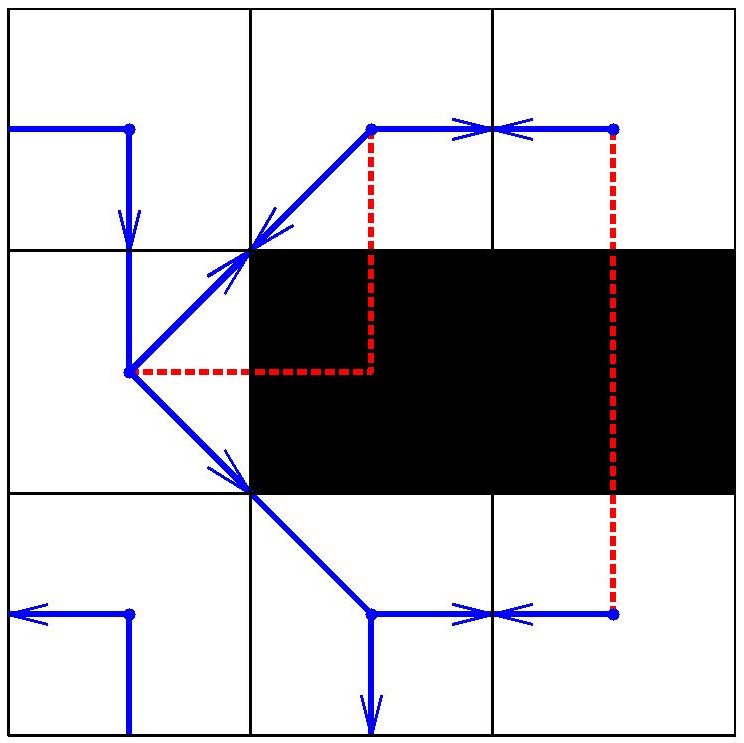}		
		} \\ 
		
		\hline
		
		%----
	
	\end{tabular}
	\caption{Strategies for $n_2 = n_1 + 3$ \\ \footnotesize{$n$ is as used in algorithm \ref{alg:HC}}}
	\label{t2}
\end{table}

\begin{table}
	\centering
	\begin{tabular}{|m{1.4cm}|m{2.9cm}|m{0.9in}|} \hline \hline		
		Node Type & Change in Path & Figure \\ 
		\hline \hline %\vspace{10em}

		%----

		\begin{tabular}{@{}c@{}} 
			 $n_1 \in \tau_1$ \\$n_2 \in \tau_2$ \\ $n_1\%16 = 0$ \\ $n_2 = n_1+1$
		\end{tabular}  
		&
		\begin{tabular}{@{}c@{}}  \\$d_1:=d(n_1+14,n_1-4)$ \\ $d_2:=d(n_1+14,n_1-2)$ \\ if($d_1<d_2$) \\$\mathcal{N}(n_1,n_1+1)$ $\leftarrow$ \\ $\mathcal{N}(n_1-4, n_1+14)$ \\ elif($d_1>d_2$) \\ $\mathcal{N}(n_1,n_1+1)$ $\leftarrow$ \\ $\mathcal{N}(n_1-2, n_1+14)$ \end{tabular}
		& 
		\parbox[c]{1.5in}{
			\includegraphics[scale=0.15, keepaspectratio]{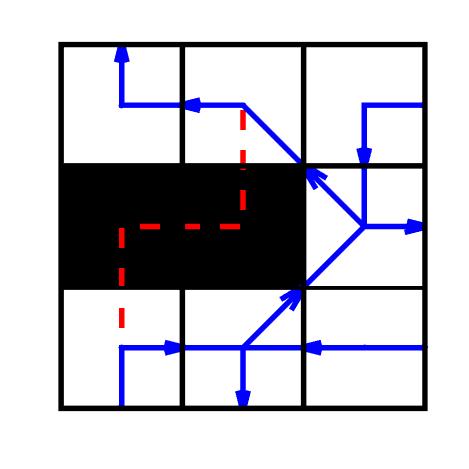}		
		} 
		\\ \hline

		\begin{tabular}{@{}c@{}} 
		$n_1 \in \tau_1$ \\$n_2 \in \tau_2$ \\ $n_1\%16 = 2$ \\ $n_2 = n_1+1$
		\end{tabular} 
		&
		\begin{tabular}{@{}c@{}}  $ \mathcal{N}(n_1,n_1+1)$ $\leftarrow$\\ $ \mathcal{N}(n_1+11, n_1+5$) \end{tabular}
		& 
		\parbox[c]{1.5in}{
			\includegraphics[scale = 0.06, keepaspectratio, angle = 270]{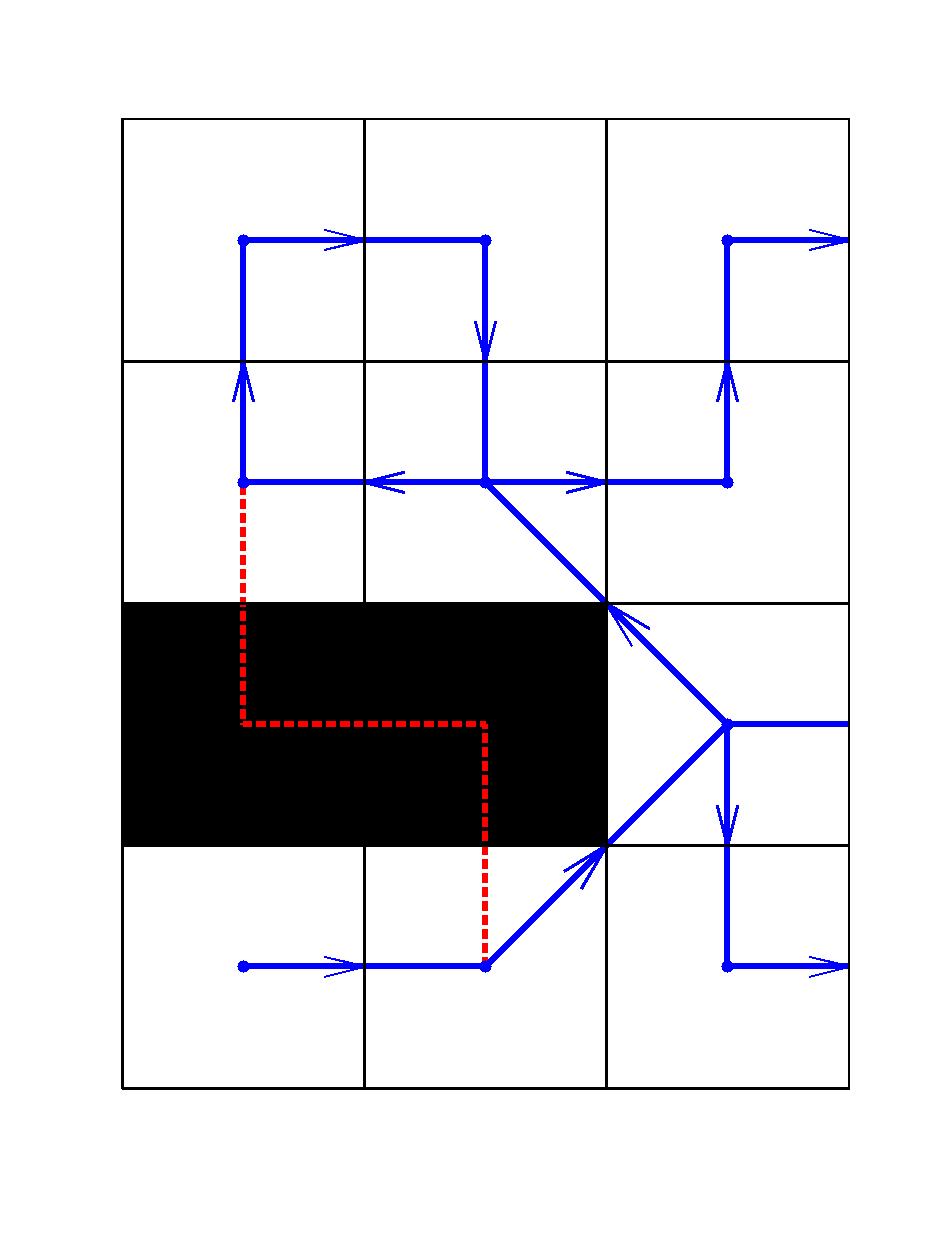}		
		} \\ 
		
		\hline

		\begin{tabular}{@{}c@{}} 
		$n_1 \in \tau_1$ \\$n_2 \in \tau_2$ \\ $n_1\%16 = 6$ \\ $n_2 = n_1+1$
		\end{tabular} 
		&
		\begin{tabular}{@{}c@{}}   $ \mathcal{N}(n_1,n_1+1)$ $\leftarrow$\\ $ \mathcal{N}(n_1-2, n_1-4)$ \end{tabular}
		& 
		\parbox[c]{1.5in}{
			\includegraphics[scale=0.06, keepaspectratio]{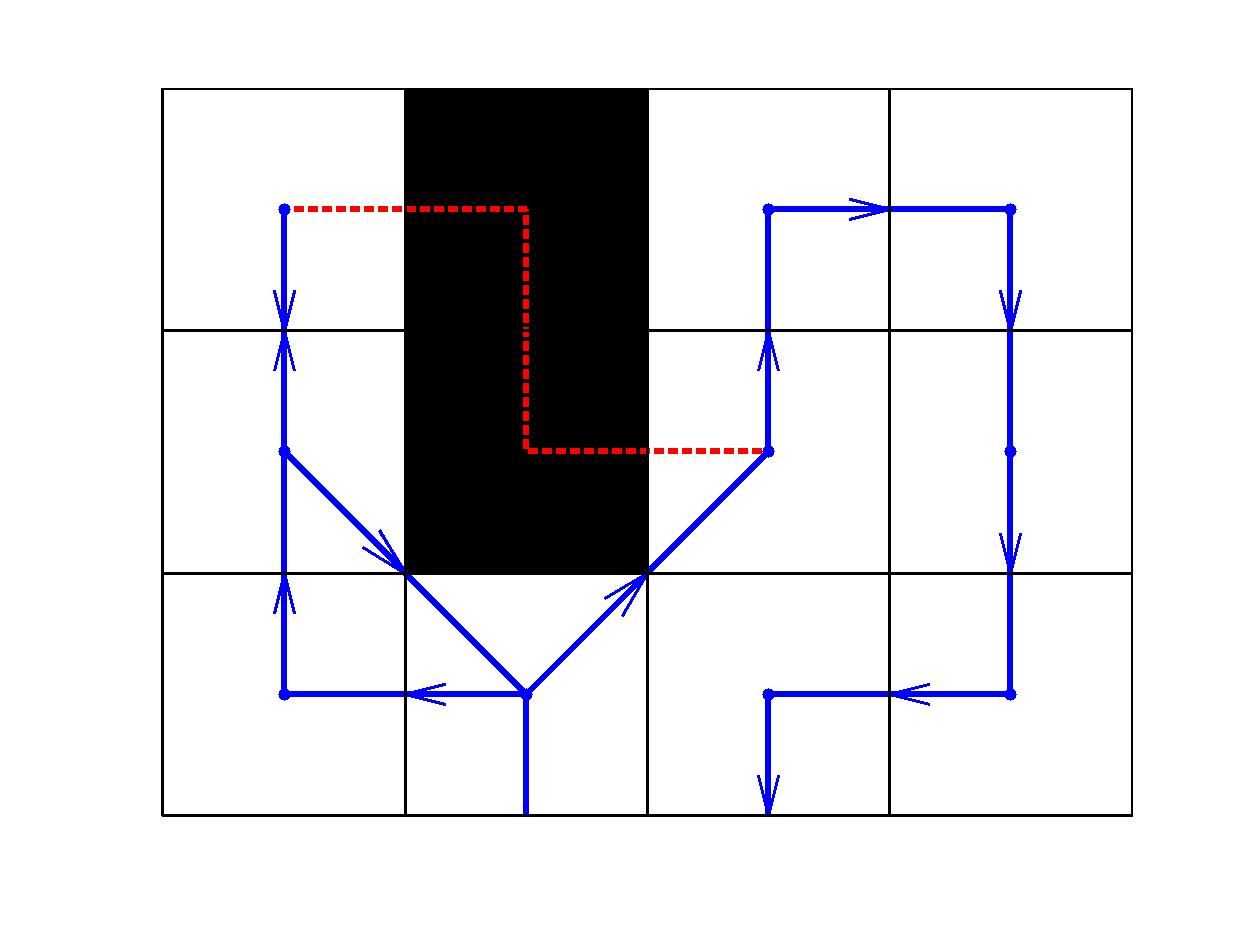}	
		} \\ 
		
		\hline

		\begin{tabular}{@{}c@{}} 
		$n_1 \in \tau_1$ \\ $n_2 \in \tau_2$ \\ $n_1\%16 = 8$ \\ $n_2 = n_1+1$
		\end{tabular} 
		&
		\begin{tabular}{@{}c@{}}  $ \mathcal{N}(n_1,n_1+1)$ $\leftarrow$ \\ $\mathcal{N}(n_1+5, n_1+3)$ \end{tabular}
		& 
		\parbox[c]{1.5in}{
			\includegraphics[scale=0.06, keepaspectratio, angle = 270]{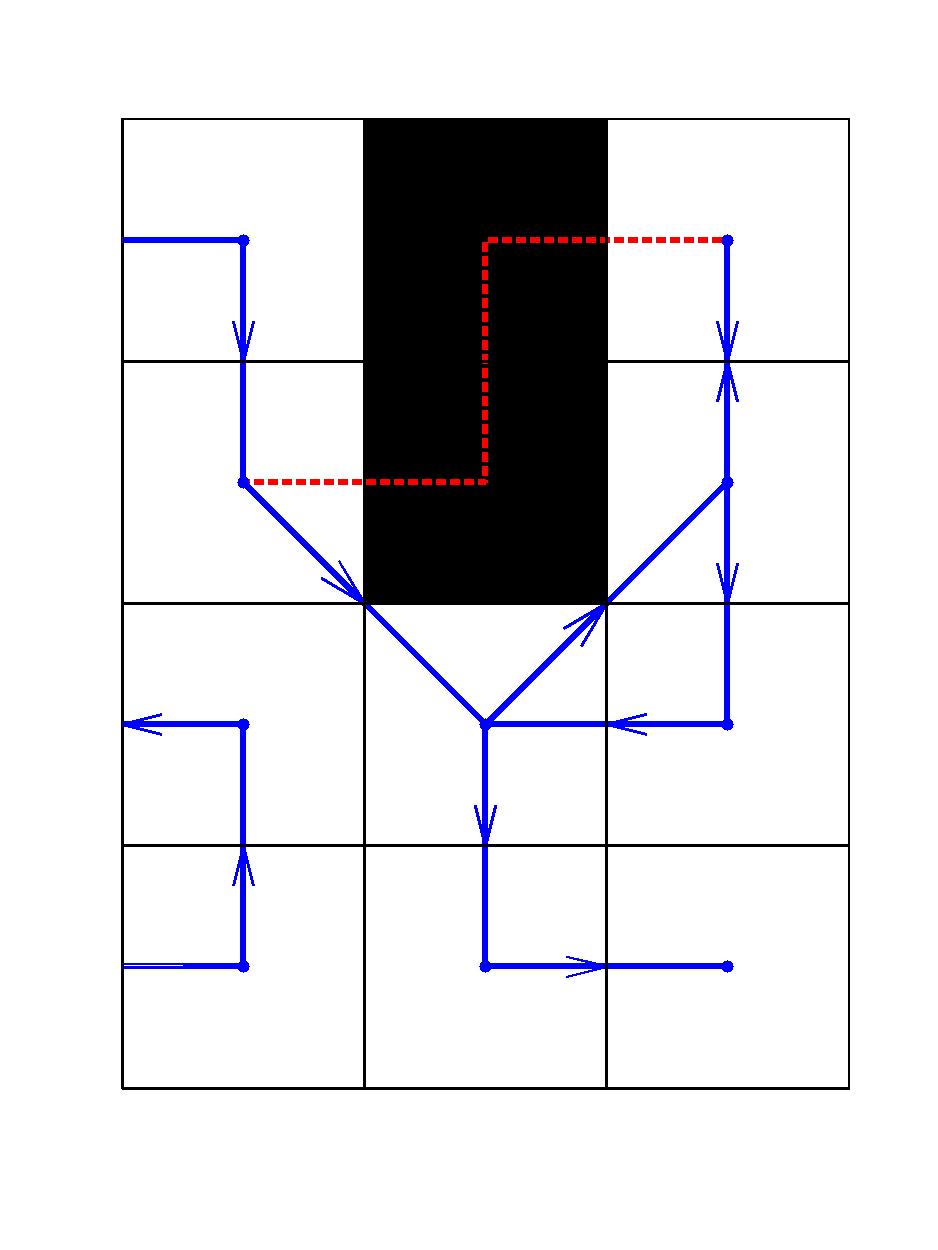}		
		} \\ 
		
		\hline

		\begin{tabular}{@{}c@{}} 
		$n_1 \in \tau_1$ \\$n_2 \in \tau_2$ \\ $n_1\%16 = 12$ \\ $n_2 = n_1+1$
		\end{tabular} 
		&
		\begin{tabular}{@{}c@{}}  $ \mathcal{N}(n_1,n_1+1)$ $\leftarrow$ \\ $ \mathcal{N}(n_1-4, n_1-10)$ \end{tabular}
		& 
		\parbox[c]{1.5in}{
			\includegraphics[scale=0.06, keepaspectratio, angle = 270]{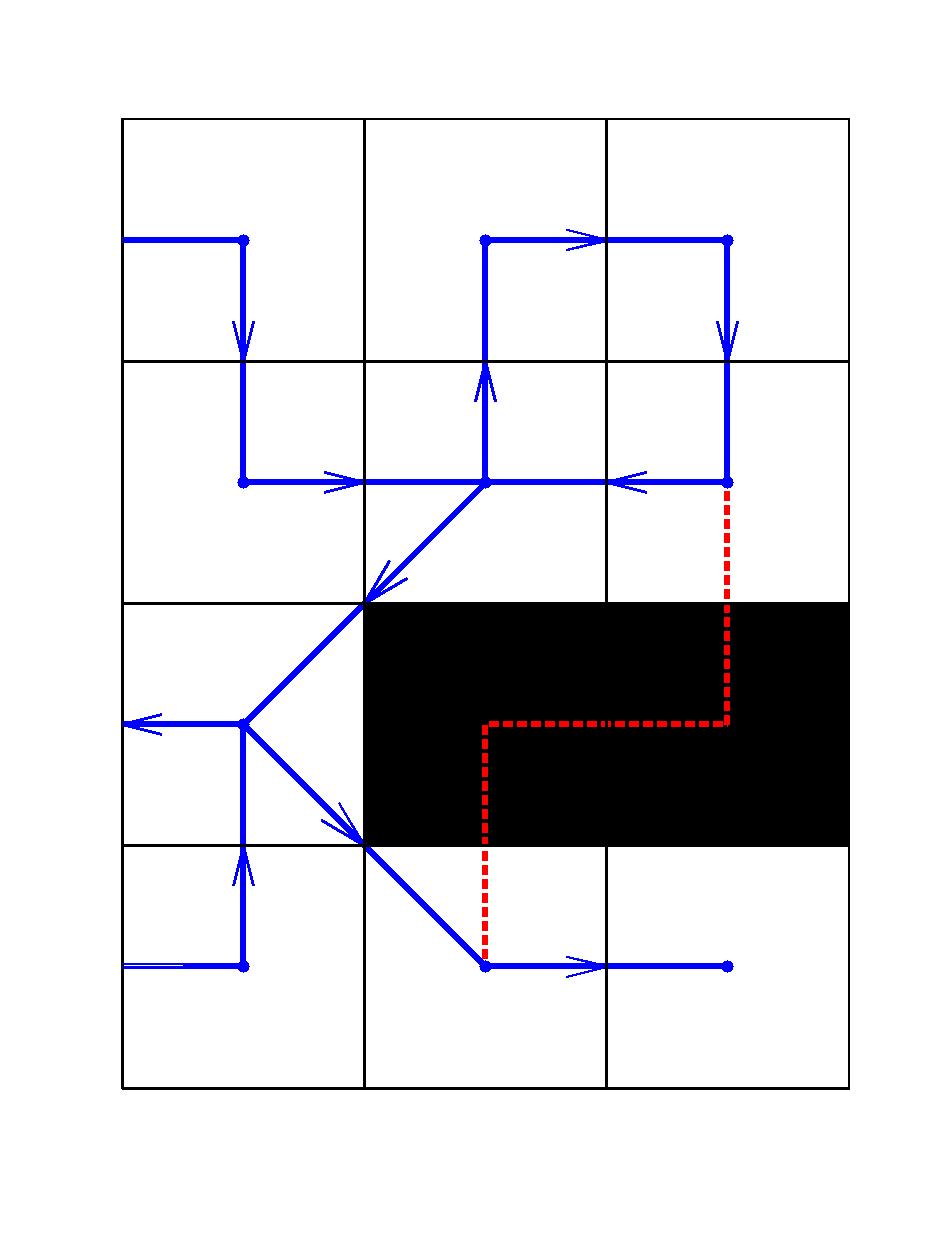}		
		} \\ 
		
		\hline

		\begin{tabular}{@{}c@{}} 
		$n_1 \in \tau_1$ \\$n_2 \in \tau_2$ \\ $n_1\%16 = 14$ \\ $n_2 = n_1+1$
		\end{tabular} 
		&
		\begin{tabular}{@{}c@{}} $d_1:=d(n_1-13,n_1+5)$ \\ $d_2:=d(n_1-13,n_1+3)$ \\ if($d_1<d_2$)\\ $\mathcal{N}(n_1,n_1+1)$ $\leftarrow$ \\ $ \mathcal{N}(n_1-13, n_1+5)$ \\ elif($d_1>d_2$) \\ $\mathcal{N}(n_1,n_1+1)$ $\leftarrow$ \\ $ \mathcal{N}(n_1-13, n_1+3)$\end{tabular}
		& 
		\parbox[c]{1.5in}{
			\includegraphics[scale=0.15, keepaspectratio]{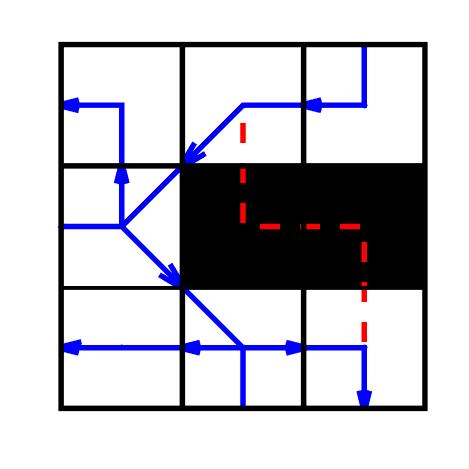}		
		} \\ 
		
		\hline

		%----
	
	\end{tabular}
	\caption{Strategies for $n_2 = n_1 + 1$}
	\label{t3}
\end{table}

\vspace{-8mm}

\bibliographystyle{IEEEtran}
\bibliography{IEEEabrv,IEEEexample}

\begin{thebibliography}{10}
\providecommand{\url}[1]{#1}
\csname url@rmstyle\endcsname
\providecommand{\newblock}{\relax}
\providecommand{\bibinfo}[2]{#2}
\providecommand\BIBentrySTDinterwordspacing{\spaceskip=0pt\relax}
\providecommand\BIBentryALTinterwordstretchfactor{4}
\providecommand\BIBentryALTinterwordspacing{\spaceskip=\fontdimen2\font plus
\BIBentryALTinterwordstretchfactor\fontdimen3\font minus
  \fontdimen4\font\relax}
\providecommand\BIBforeignlanguage[2]{{%
\expandafter\ifx\csname l@#1\endcsname\relax
\typeout{** WARNING: IEEEtran.bst: No hyphenation pattern has been}%
\typeout{** loaded for the language `#1'. Using the pattern for}%
\typeout{** the default language instead.}%
\else
\language=\csname l@#1\endcsname
\fi
#2}}

\bibitem{c5}
S.~V. Spires and S.~Y. Goldsmith, ``Exhaustive geographic search with mobile
  robots along space-filling curves,'' \emph{Collective Robotics}, pp. 1--12,
  1998.

\bibitem{sagan}
H.~Sagan, \emph{Space-Filling Curves}.\hskip 1em plus 0.5em minus 0.4em\relax
  Springer-Verlag, New York, 1994.

\bibitem{c6}
M.~Bertoldi, M.~Yardimci, C.~Pistor, and S.~Guceri, ``Domain decomposition and
  space filling curves in toolpath planning and generation,'' \emph{Solid
  Freeform Fabrication Proceedings}, pp. 267--276, 1998.

\bibitem{c7}
Z.~Liu, Y.~Chen, B.~Liu, C.~Cao, and X.~Fu, ``Hawk: An unmanned mini
  helicopter-based aerial wireless kit for localization,'' \emph{Proceedings of
  IEEE INFOCOM}, 2012.

\bibitem{c2}
B.~Goertzel, ``Global optimization with space-filling curves,'' \emph{Applied
  Mathematics Letters}, vol.~12, no.~8, pp. 133--135, 1999.

\bibitem{lpp}
C.~Gotsman and M.~Lindenbaum, ``On the metric properties of discrete
  space-filling curves,'' \emph{IEEE Transactions on Image Processing}, vol.~5,
  no.~5, pp. 794--797, May 1996.

\bibitem{bader}
M.~Bader, \emph{Space-Filling Curves: An Introduction with Applications in
  Scientific Computing}.\hskip 1em plus 0.5em minus 0.4em\relax Springer-Verlag
  Berlin Heidelberg, 2013.

\bibitem{lpp5}
K.~E. Bauman, ``The dilation factor of the peano-hilbert curve,''
  \emph{Mathematical Notes}, vol.~80, no.~5, pp. 609--620, Nov 2006.

\bibitem{lpp6}
G.~Chochia, M.~Cole, and T.~Heywood, ``Implementing the hierarchical pram on
  the 2d mesh: analyses and experiments,'' in \emph{Proceedings.Seventh IEEE
  Symposium on Parallel and Distributed Processing}, Oct 1995, pp. 587--594.

\bibitem{cluster}
B.~Moon, H.~Jagadish, C.~Faloutsos, and J.~H. Saltz, ``Analysis of the
  clustering properties of the hilbert space-filling curve,'' \emph{IEEE
  Transactions on Knowledge \& Data Engineering}, vol.~13, pp. 124--141, 01
  2001.

\bibitem{lpp7}
R.~Niedermeier, K.~Reinhardt, and P.~Sanders, ``Towards optimal locality in
  mesh-indexings,'' \emph{Discrete Applied Mathematics}, vol. 117, no.~1, pp.
  211 -- 237, 2002.

\bibitem{c9}
S.~A. Sadat, J.~Wawerla, and R.~Vaughan, ``Fractal trajectories for online
  non-uniform aerial coverage,'' \emph{Proceedings of IEEE International
  Conference on Robotics and Automation (ICRA)}, 2015.

\bibitem{pathsurvey}
E.~Galceran and M.~Carreras, ``A survey on coverage path planning for
  robotics,'' \emph{Robotics and Autonomous Systems}, vol.~61, no.~12, pp. 1258
  -- 1276, 2013.

\bibitem{sierpinski}
A.~Tiwari, H.~Chandra, J.~Yadegar, and J.~Wang, ``Constructing optimal cyclic
  tours for planar exploration and obstacle avoidance : A graph theory
  approach,'' in \emph{Advances in Cooperative Control and Optimization}.\hskip
  1em plus 0.5em minus 0.4em\relax Springer Berlin Heidelberg, 2007, pp.
  145--165.

\bibitem{SN}
S.~H. Nair, A.~Sinha, and L.~Vachhani, ``Hilbert's space-filling curve for
  regions with holes,'' in \emph{2017 IEEE 56th Annual Conference on Decision
  and Control (CDC)}, Dec 2017, pp. 313--319.

\bibitem{borkar2019aerial}
A.~V. Borkar, V.~S. Borkar, and A.~Sinha, ``Aerial monitoring of slow moving
  convoys using elliptical orbits,'' \emph{European Journal of Control},
  vol.~46, pp. 90--102, 2019.

\bibitem{expt1}
\url{https://youtu.be/G7AZaYn66xA}.

\bibitem{expt2}
\url{https://youtu.be/75qi3zcZ_dM}.

\end{thebibliography}

\end{document}